\documentclass[sigconf]{acmart}
\usepackage{todonotes}
\usepackage{subcaption}
\usepackage[most]{tcolorbox}
\usepackage{tabularray}
\usepackage{xcolor}
\UseTblrLibrary{booktabs}
\usepackage{algorithm}
\usepackage{algpseudocode}
\definecolor{algcommentblue}{RGB}{56,85,255}
\algnewcommand{\algorithmicinput}{\textbf{Input:}}
\algnewcommand{\Input}{\item[\algorithmicinput]}

\definecolor{promptbg}{RGB}{246,248,252}
\definecolor{promptbd}{RGB}{120,135,170}
\AtBeginDocument{%
  }

\setcopyright{cc}
\setcctype{by}
\copyrightyear{2026}
\acmYear{2026}
\acmDOI{10.1145/3767308.3834994}
\acmConference[MM '26]{Proceedings of the 34th ACM International Conference on Multimedia}{November 10--14, 2026}{Rio de Janeiro, Brazil}
\acmBooktitle{Proceedings of the 34th ACM International Conference on Multimedia (MM '26), November 10--14, 2026, Rio de Janeiro, Brazil}
\acmISBN{979-8-4007-2213-4/2026/11}

\usepackage{multirow}
\usepackage{colortbl}

\definecolor{level1}{RGB}{255, 249, 196}  %
\definecolor{level2}{RGB}{255, 241, 118}  %
\definecolor{level3}{RGB}{220, 231, 117}  %
\definecolor{level4}{RGB}{174, 213, 129}  %
\definecolor{level5}{RGB}{139, 195, 74}   %

\newcommand{\colorauc}[1]{%
  \ifdim #1pt < 0.6pt \cellcolor{level1!80}#1\else
  \ifdim #1pt < 0.7pt \cellcolor{level2!70}#1\else
  \ifdim #1pt < 0.8pt \cellcolor{level3!70}#1\else
  \ifdim #1pt < 0.9pt \cellcolor{level4!70}#1\else
  \cellcolor{level5!70}#1\fi\fi\fi\fi
}

\newcommand{\colormd}[1]{%
  \ifdim #1pt < 0.5pt \cellcolor{level1!80}#1\else
  \ifdim #1pt < 1.0pt \cellcolor{level2!80}#1\else
  \ifdim #1pt < 2.0pt \cellcolor{level3!80}#1\else
  \ifdim #1pt < 2.5pt \cellcolor{level4!80}#1\else
  \cellcolor{level5!80}#1\fi\fi\fi\fi
}

\newcommand{\colorsim}[1]{%
  \ifdim #1pt < 0.6pt \cellcolor{level5!70}#1\else
  \ifdim #1pt < 0.7pt \cellcolor{level4!70}#1\else
  \ifdim #1pt < 0.8pt \cellcolor{level3!70}#1\else
  \ifdim #1pt < 0.9pt \cellcolor{level2!70}#1\else
  \cellcolor{level1!80}#1\fi\fi\fi\fi
}

\begin{document}

\title{Fingerprinting Multimodal Large Language Models}

\author{Chao Huang}
\orcid{0009-0003-8876-3634}
\affiliation{%
  \institution{University of Science and Technology of China}
  \department{Anhui Province Key Laboratory of Digital Security}
  \city{Hefei}
  \state{Anhui}
  \country{China}
}
\email{chao.huang@mail.ustc.edu.cn}

\author{Meng Tong}
\orcid{0009-0009-7453-3559}
\affiliation{%
  \institution{University of Science and Technology of China}
  \department{Anhui Province Key Laboratory of Digital Security}
  \city{Hefei}
  \state{Anhui}
  \country{China}
}
\email{mtong@mail.ustc.edu.cn}

\author{Kejiang Chen}
\correspondingauthor
\orcid{0000-0002-9868-3414}
\affiliation{%
  \institution{University of Science and Technology of China}
  \department{Anhui Province Key Laboratory of Digital Security}
  \city{Hefei}
   \state{Anhui}
  \country{China}
}
\email{chenkj@ustc.edu.cn}

\renewcommand{\shortauthors}{Huang et al.}

\begin{abstract}
While multimodal large language models (MLLMs) enable a wide range of image-text reasoning tasks, recent incidents indicate that they are vulnerable to illicit deployment and unauthorized distillation. Existing solutions for model provenance are typically confounded by shared language backbones in MLLMs and struggle to detect violations of distillation. To bridge this gap and safeguard model ownership, we present the first study on multimodal model fingerprinting. Inspired by recent findings that self-attention acts as a low-pass filter and that its low-frequency components are informative, we develop AttnPrint for white-box provenance. Specifically, we extract cross-modal attention distributions and isolate their low-frequency components to serve as model fingerprints. To facilitate black-box auditing, we further introduce DistillTrace, which employs hypothesis testing of MLLM outputs to identify potential model infringement. We conduct extensive experiments on 154 model instances across 19 multimodal architectures. Notably, AttnPrint achieves strong derivative-model detection performance while remaining robust to five downstream modification techniques. DistillTrace also provides evidence of distillation relationships under three parameter-independent techniques.
\end{abstract}

\begin{CCSXML}
<ccs2012>
   <concept>
       <concept_id>10002978.10002991.10002996</concept_id>
       <concept_desc>Security and privacy~Digital rights management</concept_desc>
       <concept_significance>500</concept_significance>
       </concept>
 </ccs2012>
\end{CCSXML}

\ccsdesc[500]{Security and privacy~Digital rights management}

\keywords{multimodal model fingerprinting, copyright auditing, distillation detection}

\maketitle

\section{Introduction}

Despite the impressive performance of multimodal large language models (MLLMs) across diverse tasks~\cite{dong2024kvqa,liu2024mmc,yang2025embodiedbench}, these models increasingly face intellectual property risks, such as illicit deployment and unauthorized distillation. For instance, Anthropic publicly alleged that certain third-party models had been distilled from Claude through commercial API access.\footnote{\url{https://x.com/AnthropicAI/status/2025997928242811253?s=20}} More recently, public reports have disclosed related disputes concerning the unauthorized derivation of open-source MLLMs, including the case
involving Llama-3-V and MiniCPM-V.\footnote{\url{https://x.com/chrmanning/status/1797664513367630101}}
In such cases, attackers may apply relatively low-cost operations, such as fine-tuning, pruning, or quantization, to construct superficially modified derivative models and present them as independently developed systems. These incidents highlight the need for reliable methods to identify both modified derivatives of open-source models and students distilled from proprietary models.

A commonly used technique to safeguard model ownership is fingerprinting, wherein an auditor determines whether a suspect model was derived from a protected model through unauthorized modification or distillation. While fingerprinting has proven effective for provenance verification in LLMs~\cite{zeng2024huref,yoon2025intrinsic,zhang2025reef,wu2025gradient,pasquini2025llmmap,gao2025met,sun2025idiosyncrasies}, its direct application to multimodal large language models (MLLMs) presents substantial challenges in reliable provenance verification. Our empirical results show that text-centric fingerprints do not transfer reliably to MLLMs because they overlook cross-modal alignment. Consequently, they may conflate models with similar language backbones (see Section~\ref{sec:main-results}). Moreover, existing fingerprinting methods are largely ineffective for distillation detection, as they are primarily designed to identify direct derivation from a known base model and fail to reliably capture the behavioral relationships inherited from a specific teacher.

To bridge this gap, we present the first model fingerprinting study for reliable provenance verification of MLLMs. To address the confusion between models that share the same language backbone, we propose an attention-based fingerprinting method with white-box access to the target model. Our intuition is that differences in multimodal alignment data and training strategies lead different MLLMs to develop distinct attention patterns. In particular, cross-modal attention interactions between visual and textual tokens better capture model-specific identity information. Based on this observation, we extract cross-modal attention distributions and transform them into the frequency domain via the Fourier transform. We then retain stable low-frequency components while suppressing high-frequency noise introduced by model modifications such as fine-tuning and model merging, thereby producing a more robust fingerprint. In the black-box setting, we propose a hypothesis-testing-based method for distillation detection. Our method builds on the observation that during distillation a student model inherits behavioral characteristics of the teacher model over the distillation data and its neighboring distribution, and therefore the teacher model typically assigns higher confidence to the student model’s outputs. To reduce false positives caused by overlapping training data and inherently simple tasks, we further introduce reference models and determine whether a distillation relationship exists based on logits statistics and hypothesis testing. 

Our contributions are summarized as follows:
\begin{itemize}
\item We present the first model fingerprinting study for multimodal large language models (MLLMs).
\item We propose two methods for model copyright protection: AttnPrint uses intrinsic cross-modal attention patterns for white-box derivative-model detection, whereas DistillTrace uses reference-calibrated behavioral evidence for black-box distillation attribution.
\item We conduct large-scale evaluations on 154 model instances spanning 19 mainstream multimodal architectures. Experimental results show that AttnPrint reliably detects model derivation under five downstream modification techniques, whereas DistillTrace provides evidence of distillation relationships under three parameter-independent techniques.
\end{itemize}

\section{Preliminaries}

\subsection{Multimodal large language models}

Multimodal large language models (MLLMs) typically consist of a pretrained vision encoder, a pretrained large language model, and a connector module that bridges the two representation spaces~\cite{lin2024revolution,zhang2024mmllms}. Their training generally proceeds in two stages: multimodal pretraining on large-scale image-text pairs to achieve cross-modal alignment, followed by multimodal instruction tuning to further develop capabilities in tasks such as visual question answering, vision-language reasoning, and instruction-following dialogue~\cite{zhang2024mmllms,he2025textrich}. Because training vision encoders and large language models from scratch is prohibitively expensive, existing MLLMs often reuse the same or similar core components. For example, Qwen2-VL-7B and LLaVA-OneVision-Qwen2-7B share the Qwen2 language backbone~\cite{qwen2vl2024,llavaonevision_2024}, while many recent models also adopt ViT-style vision encoders; more examples are provided in the supplementary material. At the same time, substantial differences remain in multimodal alignment data, training strategies, and downstream fine-tuning, so even models built on similar underlying components may exhibit markedly different cross-modal interaction behaviors.

\subsection{Fingerprinting}

This paper focuses on model fingerprinting for copyright protection. Specifically, model fingerprinting aims to enable a copyright auditor to extract intrinsic characteristics from a suspicious model and compare them with the fingerprint of a victim model, in order to determine whether the suspicious model was obtained from the victim through unauthorized model modification or distillation. Model fingerprinting methods can be broadly divided into two categories: white-box fingerprinting and black-box fingerprinting.

\noindent\textbf{White-box fingerprinting.}
In the white-box auditing setting, the auditor has access not only to the input-output behavior of a model, but also to its full parameters, network architecture, intermediate representations, and gradients.
Existing methods can be grouped into three categories according to the type of internal signal they use: weight-based, activation-based, and gradient-based methods. Weight-based methods identify models by analyzing their weight parameters.
For example, HuRef treats model invariants as fingerprints and determines model attribution by comparing the similarity between invariant representations~\cite{zeng2024huref}. Similarly, a parameter-distribution-based method identifies models through statistical signatures of their weight matrices, in particular the layer-wise standard deviation patterns of attention parameters~\cite{yoon2025intrinsic}. Activation-based methods instead exploit intermediate representations produced during forward propagation. As a representative example, REEF constructs fingerprints by measuring the similarity of representation spaces across models with centered kernel alignment~\cite{zhang2025reef}. Gradient-based methods further leverage backward information. TensorGuard extracts gradient-based features and identifies model provenance through comparisons of gradient representations across models~\cite{wu2025gradient}.

\noindent\textbf{Black-box fingerprinting.}
In the black-box auditing setting, the auditor cannot access model
parameters or internal states and can only interact with the target
model through an API by observing external behaviors such as generated
texts and output probabilities. The goal is therefore to compare
response characteristics across models and infer whether a derivative
relationship exists between them. \citet{gubri2024trap} induce the
target model to produce predefined responses under carefully optimized
prompts and use the elicited outputs as model fingerprints.
\citet{pasquini2025llmmap} design discriminative query prompts and
identify model identity from differences in response style and
behavioral patterns. \citet{gao2025met} further study black-box
fingerprinting from the perspective of output distributions and
compare different models using maximum mean discrepancy (MMD).
\citet{sun2025idiosyncrasies} treat model outputs as learnable features
and train classifiers on generated texts to identify their source
models. More recently, \citet{shao2025zeroprint} approximate gradient-like responses under black-box access via zeroth-order optimization and perform source identification by comparing the resulting gradient surrogates across models.

However, existing fingerprinting methods for LLMs are primarily
designed for text-only input settings and therefore do not readily
extend to the multimodal alignment mechanism in MLLMs, which is jointly
shaped by the vision encoder, the projector, and the language model.
Our experiments show that these white-box fingerprinting methods,
originally developed for LLMs, exhibit substantially limited
identification capability in the MLLM setting. The fundamental reason
is that many MLLMs share the same LLM backbone, while existing methods
mainly characterize internal features on the language side and overlook
how visual information is encoded, aligned, and further injected into
the language generation process. As a result, when two MLLMs use the
same language backbone but differ in their vision encoders or the
datasets used for multimodal alignment, existing methods may still
misidentify them as the same model solely because of the similarity of
their language components, thereby failing to effectively capture the
cross-modal capability differences that are unique to MLLMs.

\subsection{Distillation Detection}

For distillation detection, existing methods often rely on injecting
watermarks into the teacher model in advance and exploiting their
inheritability during distillation. In particular,
\citet{gu2024learnability} study watermark distillation and show that
a student model can learn the watermark pattern carried by a
watermarked teacher's outputs. \citet{pan-etal-2025-llm} further
investigate watermark radioactivity for unauthorized knowledge
distillation and detect distillation by testing whether the student
inherits the corresponding watermark signal. However, these methods
require watermark injection before deployment and therefore rely on a
strong pre-deployment assumption. Moreover, watermark injection itself
may introduce undesirable side effects on the model's original utility
and performance.

\section{Threat Model}

In this section, we first define copyright infringement behaviors in the MLLM setting. Specifically, we consider a model owner who invests substantial resources to train an MLLM and then either releases it on an open-source platform (e.g., Hugging Face) or provides access to it as a cloud API service. We consider two representative types of infringement by an attacker. The first is unauthorized modification of open-source models. In this scenario, the attacker starts from a license-protected open-source model and applies downstream operations such as fine-tuning, pruning, quantization, or model merging to construct a derivative model that remains functionally similar to the original model while appearing superficially modified, and then claims it as an independently developed model. The second is unauthorized model distillation. This scenario may target either open-source models or closed-source models served through APIs. The attacker queries a teacher model and uses its outputs to train a student model, thereby inheriting the teacher model's capabilities.

The task of model fingerprinting is to determine whether a suspicious
model is derived from another model, which we refer to as the victim
model. In particular, this includes determining whether the suspicious
model is obtained by modifying an open-source model or by distilling
another model. We formalize this task as follows. Given a victim model
$M_v$ and a suspicious model $M_s$, the goal is to design a decision function
\begin{equation}
f(M_v, M_s) \rightarrow \{0,1\},
\end{equation}
where $f(M_v, M_s)=1$ indicates that a derivative relationship exists
between $M_s$ and $M_v$, namely that $M_s$ is obtained from $M_v$
through unauthorized modification or distillation, while
$f(M_v, M_s)=0$ indicates that no such relationship exists.

\noindent \textbf{Model owner assumptions.}
The model owner’s goal is to determine whether a suspicious model is derived from a victim model. To this end, we consider two complementary auditing settings. In the white-box setting, the auditor has access to the full parameters, network architecture, and intermediate representations of the suspicious model. In the black-box
setting, the auditor cannot access model parameters or internal states, and can only interact with the target model through an API under a limited query budget, while observing external behaviors such as generated texts, output probabilities, or logits; when necessary, the auditor may also obtain the same level of access to reference
models. For distillation detection, we further assume that the model owner can construct query samples relevant to the target task. This assumption is reasonable because distillation primarily transfers teacher-specific behavioral traits on the target task distribution, while their manifestation on unrelated tasks is typically weaker and less stable.

\section{Methods}

\begin{figure*}[t]
\centering
\includegraphics[width=\textwidth,trim=0 {\dimexpr11\baselineskip\relax} 0 2.9cm,clip]{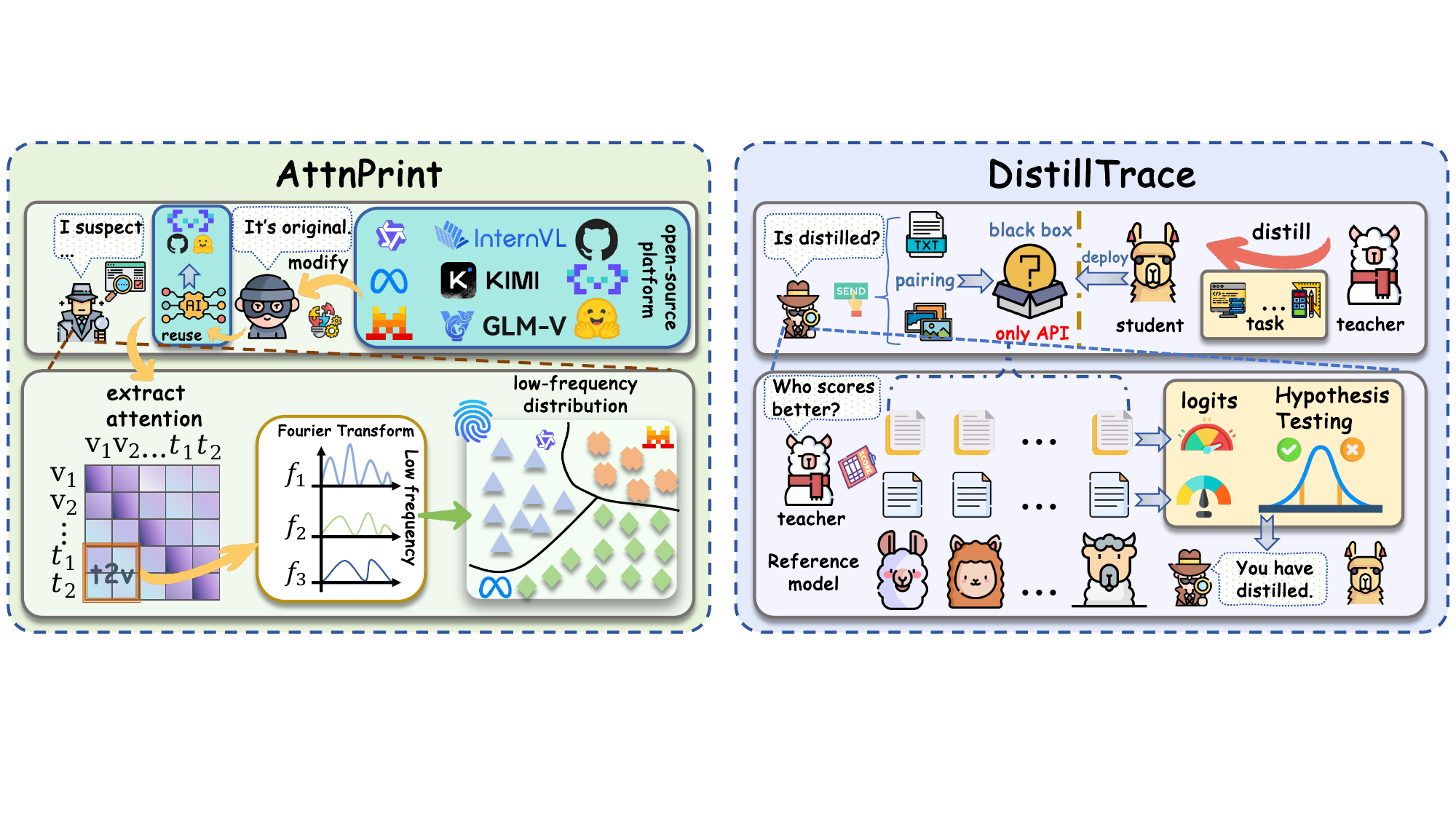}
\caption{Overview of the proposed framework.}
\label{fig:framework}
\Description{An overview of the proposed MLLM fingerprinting
framework. The white-box branch uses AttnPrint to extract frequency-
domain low-frequency cross-modal attention features for derivative-
model identification, and the black-box branch uses DistillTrace to
compare teacher confidence on suspect outputs against reference
models for distillation attribution via hypothesis testing.}
\vspace{-\baselineskip}
\end{figure*}

This section provides a detailed exposition of the two proposed methods, as illustrated in Figure~\ref{fig:framework}. For the white-box auditing scenario, we introduce \textbf{AttnPrint}, an attention-based model fingerprinting method. To address the black-box auditing scenario, we propose \textbf{DistillTrace}, which is specifically designed for the detection of knowledge distillation.

\subsection{AttnPrint}

\begin{center}
\includegraphics[width=\columnwidth]{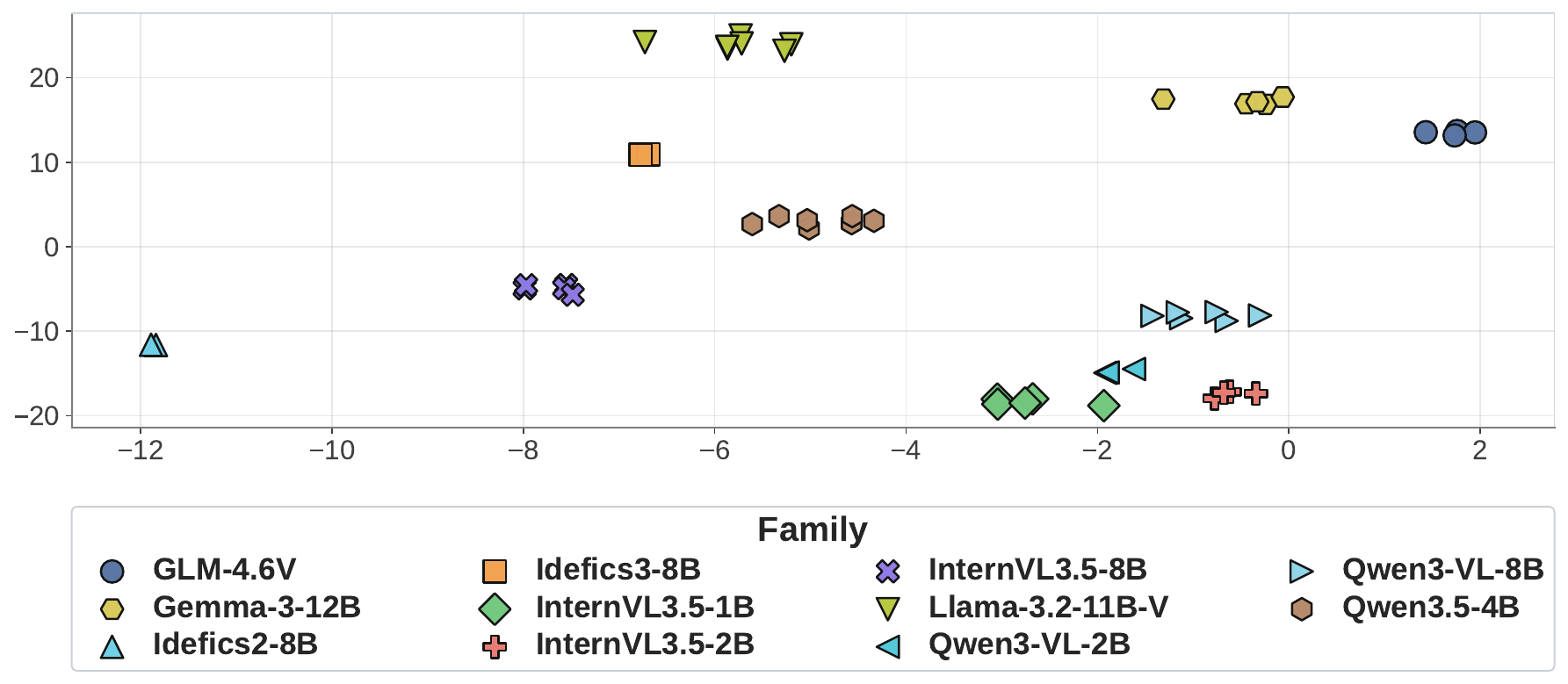}
\captionof{figure}{t-SNE visualization of attention-distribution features extracted from different MLLMs.}
\label{fig:tsne}
\Description{A two-dimensional t-SNE plot of attention-distribution
features from multiple MLLMs. Models with derivative relationships
form compact clusters, while independently trained models are more
clearly separated in the feature space.}
\end{center}

Our key insight is that differences in model architectures, training datasets, and optimization strategies shape how MLLMs allocate attention when processing the same multimodal inputs. Motivated by this insight, we examine whether attention distributions exhibit model-specific structure across MLLMs. Figure~\ref{fig:tsne} visualizes these features using t-SNE. Models with derivative relationships form compact local clusters, whereas independently trained models appear in separate regions. This exploratory visualization motivates the use of attention distributions as model fingerprints, which we evaluate quantitatively in Section~\ref{sec:main-results}.

For multimodal models, the input image and text are first encoded into visual tokens and text tokens, respectively, and then jointly participate in the subsequent cross-modal interaction and generation process within the model. Accordingly, the attention relationships in an MLLM can be further divided into four parts, namely the attention
from visual tokens to text tokens (visual-to-text), from text tokens to visual tokens (text-to-visual), from text tokens to text tokens (text-to-text), and from visual tokens to visual tokens (visual-to-visual). Among them, text-to-text attention is largely inherited from the language modeling capability of the backbone LLM, and therefore mainly reflects the intrinsic characteristics of the backbone language model itself. In contrast, text-to-visual and visual-to-text attention more directly capture the interaction mechanisms learned
during multimodal alignment, and therefore carry richer cross-modal modeling characteristics and identity information that are unique to different MLLMs. This observation is further supported by the results in the supplementary material, where text-to-visual and visual-to-text attention exhibit stronger discriminative power for model identification and fingerprinting.

Consider a model consisting of $L$ layers, where the $l$-th layer contains $H_l$ attention heads, with $T_l$ text tokens and $V_l$ visual tokens. This study focuses on the text-to-visual attention sub-matrix of the $h$-th head in the $l$-th layer, denoted as $\mathbf{A}_{l,h}^{t2v} \in \mathbb{R}^{T_l \times V_l}$ (visual-to-text attention is less
informative due to autoregressive masking; see the supplementary material). The attention vector of the $i$-th text token across all visual tokens is represented as $\mathbf{A}_{l,h}^{t2v}[i,:] \in \mathbb{R}^{V_l}$. However, post-training operations such as fine-tuning, quantization,
and model merging often perturb the attention distribution of an MLLM and introduce additional noise, thereby interfering with the stable extraction of fingerprint features. Since such noise is highly coupled with the original attention patterns, it is difficult to separate directly in the attention space. Nevertheless, existing large language models generally follow an autoregressive generation mechanism, in which the generation of each token depends on its preceding context. The resulting joint generation probability can be
written as:
\begin{equation}
p(x_{1:T}) = \prod_{t=1}^{T} p(x_t \mid x_{<t}),
\end{equation}
where the generation of each token $x_t$ depends on its preceding context $x_{<t}$. For any given position, the attention mechanism produces a set of attention weights that quantify how strongly the token at that position attends to other tokens in the context, denoted here by $\mathbf{A}_{l,h}^{t2v}[i,:]$. Following prior work that treats attention distributions over ordered token sequences as signals amenable to frequency-domain analysis~\cite{qi2026detectingcontextualhallucinationsllms}, we apply the Discrete Fourier Transform (DFT)~\cite{oppenheim2009discrete} to analyze their spectral characteristics:
\begin{equation}
X_{l,h}^{t2v}[i, k] = \sum_{n=0}^{V_l-1} \mathbf{A}_{l,h}^{t2v}[i, n] \cdot e^{-j \frac{2\pi kn}{V_l}}, \quad k = 0, 1, \dots, V_l-1,
\end{equation}
where $j=\sqrt{-1}$ denotes the imaginary unit.

We exclusively retain the low-frequency components of the signal. Intuitively, low-frequency signals reflect the global alignment logic and macroscopic processing patterns of the model, which are acquired during extensive pre-training on large-scale data distributions and thus effectively characterize the intrinsic identity of the model. In contrast, high-frequency signals are more susceptible to perturbations from downstream operations, manifesting as noise that interferes with stable fingerprint identification. This interpretation is further supported by supplementary analyses. Let $\rho \in (0, 1]$ denote the retention ratio for low-frequency components. In practice, we set $\rho = 0.2$. The ideal low-pass filter $G(k)$ is defined as:
\begin{equation}
  G(k) = \mathbf{1}\!\left(
  \min(k,V_l-k)<\frac{\rho V_l}{2}
  \right).
\end{equation}

The denoised low-frequency signal $\hat{X}_{l,h}^{t2v}[i, k]$ is defined as 
\begin{equation}
\hat{X}_{l,h}^{t2v}[i,k] = X_{l,h}^{t2v}[i,k]\cdot G(k).
\end{equation}

Finally, the signal is mapped back to the time domain through the Inverse Discrete Fourier Transform (IDFT) to yield the denoised attention vector:
\begin{equation}
\hat{A}_{l,h}^{t2v}[i,n]
=
\frac{1}{V_l}
\sum_{k=0}^{V_l-1}
\hat{X}_{l,h}^{t2v}[i,k]
\cdot
e^{j \frac{2\pi kn}{V_l}},
\quad n=0,1,\dots,V_l-1.
\end{equation}

According to Parseval's theorem~\cite{oppenheim2009discrete}, the total energy of the signal remains invariant between the time and frequency domains:
\begin{equation}
\sum_{n=0}^{V_l-1} |\hat{\mathbf{A}}[i, n]|^2 = \frac{1}{V_l} \sum_{k=0}^{V_l-1} |\hat{X}[i, k]|^2.
\end{equation}

This property allows us to examine the energy distribution of cross-modal attention in the frequency domain. Empirically, most energy is concentrated in the low-frequency band, so retaining low-frequency components preserves the dominant signal structure with limited information loss. Supporting results are provided in the supplementary material. The energy of the low-frequency attention vector for the $i$-th text token in the $h$-th head of the $l$-th layer is denoted as $E_{l,h}[i]$. To preserve token-specific attention intensity patterns, we do not average over text-token positions. Instead, we average only across the head dimension and define the layer-level fingerprint as a vector $\mathbf{S}_l \in \mathbb{R}^{T_l}$, whose $i$-th element is
\begin{equation}
S_l[i]
=
\frac{1}{H_l}\sum_{h=1}^{H_l}E_{l,h}[i]
=
\frac{1}{H_l}
\sum_{h=1}^{H_l}
\sum_{n=0}^{V_l-1}
\left|\hat{A}_{l,h}^{t2v}[i,n]\right|^2.
\end{equation}

For a model $M$ with $L$ layers, the layer-level fingerprint
$\mathbf{S}_l(M)\in\mathbb{R}^{T_l}$ may have a model-dependent
length because different models can produce different numbers of text
tokens. To obtain a fixed-dimensional representation at the layer
level, we apply mean pooling over the token dimension:
\begin{equation}
e_l(M)
=
\frac{1}{T_l}
\sum_{q=1}^{T_l} S_l(M)[q],
\qquad l=1,\ldots,L,
\end{equation}
where $e_l(M)\in\mathbb{R}$ represents the average low-frequency
cross-modal attention energy of the $l$-th layer. The model
fingerprint is then represented as the sequence of layer-wise energy
statistics:
\begin{equation}
\mathbf{F}(M)
=
\left[
e_1(M),e_2(M),\ldots,e_L(M)
\right]^{\top}
\in\mathbb{R}^{L}.
\end{equation}

Architectural heterogeneity complicates direct layer-wise fingerprint comparison, as models may differ in depth because of their original
architectures or subsequent layer pruning~\cite{men2024shortgpt,chen2024streamlining,zhang2024finercut}. We therefore use the Hungarian algorithm~\cite{kuhn1955hungarian} to align layers according to their pooled energy statistics.

Without loss of generality, let $L\leq L'$ denote the numbers of
layers in models $M$ and $M'$, respectively. We define the matching
cost between their $i$-th and $j$-th layers as
\begin{equation}
C_{i,j}(M,M')
=
\left|e_i(M)-e_j(M')\right|.
\end{equation}

We use the Hungarian algorithm to obtain a one-to-one partial matching
between the layers of the two models:
\begin{equation}
\begin{aligned}
\mathbf{X}^{\ast}
=
\arg\min_{\mathbf{X}}\quad
&\sum_{i=1}^{L}\sum_{j=1}^{L'}
C_{i,j}(M,M')x_{i,j}\\
\text{s.t.}\quad
&\sum_{j=1}^{L'}x_{i,j}=1,
&& i=1,\ldots,L,\\
&\sum_{i=1}^{L}x_{i,j}\leq1,
&& j=1,\ldots,L',\\
&x_{i,j}\in\{0,1\}.
\end{aligned}
\end{equation}
Here, $x_{i,j}=1$ indicates that the $i$-th layer of $M$ is matched
to the $j$-th layer of $M'$. Let
$\mathcal{A}^{\ast}=\{(i_k,j_k)\}_{k=1}^{K}$ denote the resulting
matched pairs, where $K=\min(L,L')$. Based on this assignment, the aligned fingerprints of the two models are defined as
\begin{equation}
\begin{aligned}
\mathbf{F}^{\ast}(M)
&=
\left[
e_{i_1}(M),e_{i_2}(M),\ldots,e_{i_K}(M)
\right]^{\top}
\in\mathbb{R}^{K},\\
\mathbf{F}^{\ast}(M')
&=
\left[
e_{j_1}(M'),e_{j_2}(M'),\ldots,e_{j_K}(M')
\right]^{\top}
\in\mathbb{R}^{K}.
\end{aligned}
\end{equation}

We then measure the agreement between the two aligned fingerprints
using the Pearson correlation coefficient:
\begin{equation}
r(M,M')
=
\operatorname{Corr}
\left(
\mathbf{F}^{\ast}(M),
\mathbf{F}^{\ast}(M')
\right),
\end{equation}
where $\operatorname{Corr}(\cdot,\cdot)$ denotes the Pearson correlation coefficient. A higher $r(M,M')$ indicates stronger agreement between the matched layer-wise attention-energy patterns and thus stronger evidence of a potential derivative relationship.

\subsection{DistillTrace}

Knowledge distillation typically utilizes the Kullback-Leibler (KL) divergence to measure the discrepancy between the soft distribution produced by a student model $P_S^T$ and that of a teacher model $P_T^T$~\cite{hinton2015distilling}:
\begin{equation}
\mathcal{L}_{\mathrm{KD}} = T^2 \sum_i p_i^{(T)} \log \left(\frac{p_i^{(T)}}{q_i^{(T)}}\right).
\end{equation}

This implies that, if a student model is distilled from a teacher model, its generation behavior on the distillation task distribution should better fit the teacher model's output preference. As a result, outputs generated by the student model are expected to receive higher confidence under the teacher model. Our goal is therefore to determine whether the teacher model assigns higher logits to a given output sequence $\mathbf{y} = \{y_1, y_2, \dots, y_N\}$ generated by a student model. Since commercial model APIs typically do not expose raw logits, we instead use a monotonic log-odds score derived from the returned log-probabilities. For a token $y_t$ with probability $p_t = \exp(\text{logprob}_t)$, the log-odds confidence score is calculated as:
\begin{equation}
\text{Logit}(p_t) = \log \left( \frac{p_t}{1 - p_t} \right).
\end{equation}

We then use the average logit over the entire sequence to measure its overall confidence, thereby avoiding unfairly lower scores for longer sequences caused by probability accumulation.
\begin{equation}
\bar{L}(M, \mathbf{y}^{(i)}) = \frac{1}{N} \sum_{t=1}^{N} \text{Logit}(p_t \mid y_{<t}, M).
\end{equation}
To rigorously determine whether a distillation relationship exists between a suspect model and a candidate teacher model, we formulate the problem as a hypothesis testing task based on paired score differences. Given a suspect model $M_S$, a candidate teacher model $M_T$, and $b$ independent reference models $\{M_{R,1},
M_{R,2}, \dots, M_{R,b}\}$, for the $i$-th query, we first compute the average logit assigned by the teacher model to the output
generated by the suspect model:
\begin{equation}
\bar{L}(M_T, \mathbf{y}_S^{(i)}),
\end{equation}
where $\mathbf{y}_S^{(i)}$ denotes the output sequence produced by the suspect model $M_S$ under the $i$-th query. We then let each reference model generate an output $\mathbf{y}_{R,j}^{(i)}$ under the same
query, and compute the average logit assigned by the teacher model to these reference outputs. Based on these scores, we construct the reference baseline as
\begin{equation}
\bar{L}_{\mathrm{refs}}^{(i)}
=
\frac{1}{b}\sum_{j=1}^{b}\bar{L}(M_T, \mathbf{y}_{R,j}^{(i)}).
\end{equation}

The introduction of reference models serves two purposes: first, to approximate the output distribution of independent models; and second, to calibrate for sample difficulty. Without such calibration, high
teacher confidence on a given sample may simply reflect that the sample is intrinsically easy for most models, rather than that the suspect model preserves teacher-specific decision characteristics.

On this basis, we define the paired difference for the $i$-th query as
\begin{equation}
\Delta_i
=
\bar{L}(M_T,\mathbf{y}_S^{(i)})
-
\bar{L}_{\mathrm{refs}}^{(i)}.
\end{equation}

If the suspect model is indeed distilled from the teacher model, its generation behavior should better align with the teacher model’s output preference, and the distribution of the paired differences is therefore expected to exhibit a positive location shift. To test whether this effect is statistically significant, we apply a one-sided Wilcoxon signed-rank test to the paired differences $\{\Delta_i\}_{i=1}^{n}$, assuming that they are independent and drawn from an approximately symmetric distribution. Specifically, we define the null hypothesis $H_0$ and the alternative hypothesis $H_1$ as follows:
\begin{equation}
\begin{aligned}
H_0:\theta_{\Delta}\leq 0,
\\
H_1:\theta_{\Delta}>0,
\end{aligned}
\end{equation}
where $\theta_{\Delta}$ denotes the location-shift parameter of the paired-difference distribution.

Based on the paired differences, we apply the Wilcoxon signed-rank test. After removing zero differences, we rank the absolute values of the remaining differences:
\begin{equation}
R_j = \operatorname{rank}(|\Delta_j|).
\end{equation}

Based on these ranks, the Wilcoxon signed-rank statistic is defined as
\begin{equation}
W^+ = \sum_{\Delta_j > 0} R_j,
\end{equation}
which is the sum of the ranks corresponding to all positive
differences. Intuitively, if the suspect model better matches the
teacher model's preference than the reference models, positive
differences should not only occur more frequently, but also tend to
have larger magnitudes, leading to a larger $W^+$.

The $p$-value is then defined as the probability, under the null
hypothesis, of observing a positive rank sum at least as large as the one obtained:
\begin{equation}
p = P\!\left(W^+_{\mathrm{null}} \ge W^+_{\mathrm{obs}}\right).
\end{equation}

A smaller $p$-value indicates that the observed positive shift is less
likely to arise from random variation, and therefore provides stronger
evidence for the existence of a distillation relationship between the
suspect model and the candidate teacher model.

\section{Experiments}

\subsection{Experimental Setup}

Our experimental setup largely follows \textsc{LEAFBENCH}~\cite{shao2025sok}, a benchmark
  designed for model fingerprinting evaluation. Since
\textsc{LEAFBENCH} is originally designed for text-only LLMs, we replace the
LLMs in its benchmark with MLLMs to accommodate the multimodal
setting, while retaining its original configurations for model
modification operations and evaluation protocols.

\noindent \textbf{Models.} 
We evaluate representative MLLMs from the Qwen, InternVL, Llama,
Gemma, LLaVA, GLM, Kimi-VL, Pixtral and Idefics families. The Qwen models are
Qwen3.5-4B~\cite{qwen3.5}, Qwen3-VL-8B, Qwen3-VL-2B~\cite{qwen3vl2025},
Qwen2-VL-7B, Qwen2-VL-2B~\cite{qwen2vl2024}, and Qwen2.5-VL-7B~\cite{qwen25vl2025}. The InternVL models are InternVL3.5-8B, InternVL3.5-2B, and InternVL3.5-1B~\cite{internvl35_2025}. The Idefics models are Idefics2-8B~\cite{laurençon2024mattersbuildingvisionlanguagemodels} and Idefics3-8B-Llama3~\cite{laurençon2024buildingbetterunderstandingvisionlanguage}. The remaining models are
Llama-3.2-11B-Vision~\cite{llama32vision_2024}, Gemma-3-12B~\cite{gemma3_2025} and Gemma-3-4B~\cite{gemma3_2025}, LLaVA-v1.6-Mistral-7B~\cite{llava16_2024}, and LLaVA-OneVision-Qwen2-7B~\cite{llavaonevision_2024}. We further include GLM-4.6V-Flash~\cite{glm46v_2025}, Kimi-VL-A3B-Instruct~\cite{kimivl_2025}, and Pixtral\nobreakdash-12B~\cite{pixtral12b_2024}. Overall, these models range
from 1B to 16B parameters, and we apply the parameter-altering techniques described below to construct 154 model instances.

\noindent \textbf{Baselines.} We compare our methods with seven fingerprinting baselines originally developed for LLMs. The four white-box baselines are HuRef~\cite{zeng2024huref}, PDF~\cite{yoon2025intrinsic}, REEF~\cite{zhang2025reef}, and TensorGuard~\cite{wu2025gradient}, while the three black-box baselines are LLMmap~\cite{pasquini2025llmmap}, MET~\cite{gao2025met}, and SEF~\cite{shao2025sok}.

\noindent \textbf{Metrics.} 
We use four metrics to evaluate the identification performance of
model fingerprinting methods: Area Under the ROC Curve (AUC), accuracy (ACC), TPR@1\%FPR, and Mahalanobis Distance (MD)~\cite{mahalanobis1936generalized}. AUC is a threshold-independent metric that measures the overall ability of a method to distinguish derived models from independent models. ACC measures the overall classification accuracy under the selected decision threshold. TPR@1\%FPR reflects the detection capability of a method under a low-false-positive regime. Throughout the paper, TPR@1\%FPR is reported in decimal form. MD measures the separability between derived and independent models
in fingerprint scores. A higher MD indicates a clearer decision boundary and more stable fingerprints under subtle perturbations.

\noindent \textbf{Techniques affecting model fingerprinting.}
After pre-training, models often undergo a series of downstream modifications before deployment. We categorize these modifications into two groups according to whether they directly alter model parameters: parameter-altering techniques and parameter-independent techniques.

Parameter-altering techniques refer to modifications that directly change model parameters, including full fine-tuning (FT)~\cite{devlin2019bert}, parameter\nobreakdash-efficient fine-tuning
(PEFT)~\cite{hu2021lora}, quantization (QZ)~\cite{frantar2022gptq}, model
merging (MM)~\cite{wortsman2022modelsoups}, pruning (PR)~\cite{sun2024wanda,men2024shortgpt},
and distillation~\cite{hinton2015distilling}. For the base models and their FT, PEFT, QZ, and MM variants, we use 75 publicly available checkpoints from Hugging Face; the corresponding model IDs are provided in the supplementary material. For pruning, we construct 15 pruned  
models using both unstructured pruning with
Wanda~\cite{sun2024wanda} and structured pruning with ShortGPT~\cite{men2024shortgpt}. For distillation, we use GLM-4.6V-Flash, Qwen3-VL-8B-Instruct, Gemma-3-12B-It, and Pixtral\nobreakdash-12B as teacher models; GeomVerse~\cite{kazemi2023geomverse}, Localized Narratives~\cite{pont2020localized}, TabMWP~\cite{lu2022tabmwp}, and
WebSight~\cite{laurencon2024websight} as distillation datasets; and Gemma-3-4B-Instruct, InternVL3.5-1B-hf, InternVL3.5\nobreakdash-2B-hf, and Qwen2-VL-2B-Instruct as
student models, resulting in 64 distilled models. Detailed pruning and distillation settings are provided in the supplementary material.

Parameter-independent techniques do not change model parameters, but
instead affect model behavior at inference time. In this category, we consider system prompts (SP)~\cite{wallace2024instructionhierarchy}, sampling
strategies (SS)~\cite{fan2018hierarchical,holtzman2020curious}, and
retrieval-augmented generation
(RAG)~\cite{lewis2020retrieval}. Detailed configurations of these settings are provided in the supplementary material.

\vspace{-0.3em}
\subsection{Main Results}
\label{sec:main-results}

\begin{table*}[t]
\centering
\tabcolsep=1.1mm
\renewcommand{\arraystretch}{1.2}
\caption{Overall performance comparison of AttnPrint and baselines under model modification scenarios.}
\label{tab:overall}
\scalebox{0.82}{
\begin{tabular}{l c c c c c c c c}
\toprule
 & \multicolumn{5}{c}{\textbf{White-box}} & \multicolumn{3}{c}{\textbf{Black-box}} \\
\cmidrule(lr){2-6} \cmidrule(lr){7-9}
\textbf{Metric} & HuRef & PDF & REEF & TensorGuard & AttnPrint & LLMmap & MET & SEF \\
\midrule
\textbf{AUC} $\uparrow$ & 0.9777($\pm$0.012) & 0.9309($\pm$0.015) & 0.8914($\pm$0.011) & 0.8464($\pm$0.018) & \textbf{0.9937}($\pm$0.006) & 0.7731($\pm$0.017) & 0.5638($\pm$0.014) & 0.7719($\pm$0.016)\\
\textbf{ACC} $\uparrow$ & 0.9861($\pm$0.009) & 0.9800($\pm$0.011) & 0.8840($\pm$0.014) & 0.8889($\pm$0.013) & \textbf{0.9879}($\pm$0.008) & 0.8849($\pm$0.015) & 0.5515($\pm$0.019) & 0.8863($\pm$0.012)\\
\textbf{TPR@1\%FPR} $\uparrow$ & 0.8632($\pm$0.021) & 0.7863($\pm$0.024) & 0.6838($\pm$0.018) & 0.4884($\pm$0.022) & \textbf{0.9636}($\pm$0.017) & 0.3189($\pm$0.020) & 0.0000($\pm$0.000) & 0.3428($\pm$0.019)\\
\textbf{MD} $\uparrow$ & 2.5656($\pm$0.063) & 2.6364($\pm$0.057) & 2.9369($\pm$0.049) & 0.8787($\pm$0.036) & \textbf{3.4428}($\pm$0.068) & 0.7241($\pm$0.031) & 0.2560($\pm$0.018) & 1.1406($\pm$0.042)\\
\bottomrule
\end{tabular}
}
\end{table*}

\textbf{Overall performance.} Table~\ref{tab:overall} compares AttnPrint with the fingerprinting baselines. AttnPrint achieves state-of-the-art results across all metrics, demonstrating the strongest overall identification performance. Among the white-box baselines, HuRef is the most competitive, but still falls short of AttnPrint, especially at low false positive rates. In contrast, black-box methods perform substantially worse across all metrics, due to their lack of access to internal model information. These results show that exploiting cross-modal attention signals provides a more effective fingerprint than existing white-box baselines, while access to internal model information remains crucial for reliable MLLM fingerprinting.

\begin{table}[h]
\centering
\tabcolsep=1.0mm
\renewcommand{\arraystretch}{1.1}
\caption{Performance comparison of model fingerprinting methods across different parameter-altering techniques.}
\label{tab:pa}
\scalebox{0.81}{
\begin{tabular}{lll ccccc}
\toprule
\textbf{Type} & \textbf{Method} & \textbf{Metric} & \textbf{FT} & \textbf{PEFT} & \textbf{QZ} & \textbf{MM} & \textbf{PR} \\
\midrule
\multirow{15}{*}{\textbf{White-box}} & \multirow{3}{*}{HuRef}
  & AUC $\uparrow$      & \colorauc{0.998} & \colorauc{1.000} & \colorauc{1.000} & \colorauc{1.000} & \colorauc{0.996} \\
  & & TPR@1\%FPR $\uparrow$     & \colorauc{0.868} & \colorauc{1.000} & \colorauc{1.000} & \colorauc{1.000} & \colorauc{0.851} \\
  & & MD $\uparrow$       & \colormd{2.928} & \colormd{3.241} & \colormd{3.404} & \colormd{3.180} & \colormd{3.067} \\
\cmidrule{2-8}

& \multirow{3}{*}{PDF}
  & AUC $\uparrow$      & \colorauc{0.943} & \colorauc{1.000} & \colorauc{0.887} & \colorauc{0.842} & \colorauc{0.908} \\
  & & TPR@1\%FPR $\uparrow$     & \colorauc{0.763} & \colorauc{1.000} & \colorauc{0.833} & \colorauc{0.750} & \colorauc{0.714} \\
  & & MD $\uparrow$       & \colormd{2.597} & \colormd{2.741} & \colormd{2.724} & \colormd{2.694} & \colormd{2.668} \\
\cmidrule{2-8}

& \multirow{3}{*}{REEF}
  & AUC $\uparrow$      & \colorauc{0.823} & \colorauc{0.941} & \colorauc{0.769} & \colorauc{0.958} & \colorauc{0.891} \\
  & & TPR@1\%FPR $\uparrow$     & \colorauc{0.579} & \colorauc{0.833} & \colorauc{0.417} & \colorauc{0.750} & \colorauc{0.625} \\
  & & MD $\uparrow$       & \colormd{2.727} & \colormd{3.422} & \colormd{1.858} & \colormd{2.868} & \colormd{2.954} \\
\cmidrule{2-8}

& \multirow{3}{*}{TensorGuard}
  & AUC $\uparrow$      & \colorauc{0.835} & \colorauc{0.989} & \colorauc{0.767} & \colorauc{0.686} & \colorauc{0.843} \\
  & & TPR@1\%FPR $\uparrow$     & \colorauc{0.414} & \colorauc{0.750} & \colorauc{0.182} & \colorauc{0.333} & \colorauc{0.400} \\
  & & MD $\uparrow$       & \colormd{0.836} & \colormd{1.217} & \colormd{0.604} & \colormd{0.459} & \colormd{0.894} \\
\cmidrule{2-8}

& \multirow{3}{*}{AttnPrint}
  & AUC $\uparrow$      & \colorauc{0.999} & \colorauc{1.000} & \colorauc{1.000} & \colorauc{1.000} & \colorauc{0.997} \\
  & & TPR@1\%FPR $\uparrow$     & \colorauc{0.952} & \colorauc{1.000} & \colorauc{1.000} & \colorauc{1.000} & \colorauc{0.941} \\
  & & MD $\uparrow$       & \colormd{3.363} & \colormd{3.266} & \colormd{3.763} & \colormd{3.671} & \colormd{3.588} \\
\midrule

\multirow{9}{*}{\textbf{Black-box}}
& \multirow{3}{*}{LLMmap}
  & AUC $\uparrow$      & \colorauc{0.706} & \colorauc{0.770} & \colorauc{0.739} & \colorauc{0.651} & \colorauc{0.724} \\
  & & TPR@1\%FPR $\uparrow$     & \colorauc{0.342} & \colorauc{0.333} & \colorauc{0.333} & \colorauc{0.250} & \colorauc{0.319} \\
  & & MD $\uparrow$       & \colormd{0.304} & \colormd{0.700} & \colormd{0.803} & \colormd{0.531} & \colormd{0.614} \\
\cmidrule{2-8}

& \multirow{3}{*}{MET}
  & AUC $\uparrow$      & \colorauc{0.559} & \colorauc{0.509} & \colorauc{0.495} & \colorauc{0.454} & \colorauc{0.528} \\
  & & TPR@1\%FPR $\uparrow$     & \colorauc{0.000} & \colorauc{0.000} & \colorauc{0.000} & \colorauc{0.000} & \colorauc{0.000} \\
  & & MD $\uparrow$       & \colormd{0.239} & \colormd{0.037} & \colormd{0.024} & \colormd{0.226} & \colormd{0.121} \\
\cmidrule{2-8}

& \multirow{3}{*}{SEF}
  & AUC $\uparrow$      & \colorauc{0.827} & \colorauc{0.937} & \colorauc{0.750} & \colorauc{0.833} & \colorauc{0.872} \\
  & & TPR@1\%FPR $\uparrow$     & \colorauc{0.556} & \colorauc{0.800} & \colorauc{0.500} & \colorauc{0.500} & \colorauc{0.589} \\
  & & MD $\uparrow$       & \colormd{1.277} & \colormd{1.720} & \colormd{0.905} & \colormd{1.172} & \colormd{1.341} \\
\bottomrule
\end{tabular}
}
\vspace{-1\baselineskip}
\end{table}

\noindent\textbf{Parameter-altering modifications.}
Table~\ref{tab:pa} shows that AttnPrint achieves the best or tied-best
AUC and TPR@1\%FPR across all five modifications and the highest MD
in most settings. The advantage is particularly clear under
quantization and model merging, where several representation- and
gradient-based baselines degrade substantially.

\begin{table*}[t]
\centering
\tabcolsep=1.1mm
\renewcommand{\arraystretch}{1.2}
\caption{Cross-family similarity comparison on selected model pairs from the benchmark configuration. Lower similarity indicates better model
separation.}
\label{tab:cross-family-similarity}
\scalebox{0.80}{
\begin{tabular}{l c c c c c c c c}
\toprule
  & \multicolumn{5}{c}{\textbf{White-box}} & \multicolumn{3}{c}{\textbf{Black-box}} \\
\cmidrule(lr){2-6} \cmidrule(lr){7-9}
\textbf{Similarity} $\downarrow$ & HuRef & PDF & REEF & TensorGuard & AttnPrint & LLMmap & MET & SEF \\
\midrule
\textbf{Qwen3-VL-8B / InternVL3.5-8B} & \colorsim{0.9688} & \colorsim{0.9969} & \colorsim{0.5593} & \colorsim{0.9970} & \colorsim{0.6530} &
\colorsim{0.9990} & \colorsim{0.9950} & \colorsim{0.9570}\\
\textbf{InternVL3.5-2B / Qwen3-VL-2B} & \colorsim{0.9844} & \colorsim{0.9985} & \colorsim{0.5059} & \colorsim{0.9964} & \colorsim{0.4309} &
\colorsim{0.9981} & \colorsim{0.9921} & \colorsim{0.9510}\\
\textbf{Idefics2-8B / Pixtral-12B} & \colorsim{0.9720} & \colorsim{0.9948} & \colorsim{0.3180} & \colorsim{0.9962} & \colorsim{0.2410} &
\colorsim{0.9987} & \colorsim{0.9942} & \colorsim{0.9560}\\
\textbf{Pixtral-12B / LLaVA-v1.6-Mistral-7B} & \colorsim{0.7109} & \colorsim{0.9634} & \colorsim{0.5401} & \colorsim{0.9886} & \colorsim{0.5219}
& \colorsim{0.9992} & \colorsim{0.9935} & \colorsim{0.9520}\\
\textbf{LLaVA-v1.6-Mistral-7B / Idefics2-8B} & \colorsim{0.9815} & \colorsim{0.9976} & \colorsim{0.2860} & \colorsim{0.9995} & \colorsim{0.2190}
& \colorsim{0.9989} & \colorsim{0.9946} & \colorsim{0.9490}\\
\bottomrule
\end{tabular}
}
\end{table*}

\noindent \textbf{Cross-Family Pairs with Similar Language Backbones.}
Table~\ref{tab:cross-family-similarity} compares five selected cross-family model pairs whose language backbones are the same or highly similar. Specifically, Qwen3-VL-8B and InternVL3.5-8B are both built on the Qwen3-8B language backbone; InternVL3.5-2B and Qwen3-VL-2B both use Qwen3-2B; Idefics2-8B and LLaVA-v1.6-Mistral-7B are based on Mistral-7B; and Pixtral-12B uses a 12B variant from the Mistral series. Since lower similarity indicates better separation, this experiment tests whether a fingerprinting method can avoid confusing independently trained MLLMs despite shared language backbones. Most baselines still assign uniformly high similarity to these pairs. In particular, PDF, TensorGuard, HuRef and the black-box methods remain highly similar on almost all pairs, indicating limited ability to separate MLLMs that share the same language backbone. REEF and AttnPrint are the only methods that substantially reduce similarity on these hard pairs. Among them, AttnPrint achieves the lowest similarity on four of the five pairs and the second-lowest on the remaining pair (Qwen3-VL-8B / InternVL3.5-8B), where REEF attains 0.5593 and AttnPrint attains 0.6530. This strong performance of AttnPrint suggests that cross-modal attention better captures the multimodal differences among MLLMs with shared language backbones. REEF also performs well on these pairs, suggesting that representation-space features capture multimodal alignment shifts. However, its lower overall AUC indicates that it may over-separate same-family models, reducing overall reliability.

\begin{table}[t]
\centering
\tabcolsep=0.8mm
\renewcommand{\arraystretch}{1.2}
\caption{Performance comparison of different model fingerprinting methods on distilled models.}
\label{tab:distill}
\scalebox{0.73}{
\begin{tabular}{l c c c c c c c c}
\toprule
 & \multicolumn{4}{c}{\textbf{White-box}} & \multicolumn{4}{c}{\textbf{Black-box}} \\
\cmidrule(lr){2-5} \cmidrule(lr){6-9}
\textbf{Metric} & HuRef & PDF & REEF & TensorGuard & LLMmap & MET & SEF & DistillTrace \\
\midrule
\textbf{AUC} $\uparrow$ & 0.4942 & 0.5000 & 0.5135 & 0.4941 & 0.5000 & 0.5000 & 0.5010 & \textbf{0.8507}\\
\textbf{TPR@1\%FPR} $\uparrow$ & 0.0000 & 0.0000 & 0.0000 & 0.0000 & 0.0179 & 0.0000 & 0.0000 & \textbf{0.3281}\\
\bottomrule
\end{tabular}
}
\end{table}

\noindent \textbf{Distillation Detection.} By default, DistillTrace is evaluated with 200 queries and four reference models. Table~\ref{tab:distill} shows that existing white-box and black-box fingerprinting baselines achieve near-random performance, whereas DistillTrace achieves an AUC of 0.8507 and a TPR@1\%FPR of 0.3281. This advantage can be attributed to the fact that distillation causes only limited changes to the student’s fingerprint and primarily preserves relative alignment with the teacher, rather than absolute fingerprint similarity.
Conventional fingerprinting methods mainly capture such absolute alignment and therefore struggle to identify distillation relationships. By introducing reference models, DistillTrace measures the suspect model’s relative alignment with the candidate teacher and isolates teacher-specific characteristics.

\begin{table}[t]
\centering
\tabcolsep=7mm
\renewcommand{\arraystretch}{1.1}
\caption{DistillTrace performance for distillation detection under different parameter-independent techniques.}
\label{tab:distill_variant}
\scalebox{0.84}{
\begin{tabular}{lccc}
\toprule
\textbf{Metric} & \textbf{SP} & \textbf{RAG} & \textbf{SS} \\
\midrule
AUC $\uparrow$ & 0.7035 & 0.7936 & 0.7136 \\
TPR@1\%FPR $\uparrow$ & 0.1233 & 0.1850 & 0.1033 \\
\bottomrule
\end{tabular}
}
\vspace{-\baselineskip}
\end{table}

\noindent \textbf{Distillation Detection under Parameter-Independent Techniques.} Table~\ref{tab:distill_variant} shows that DistillTrace degrades under parameter-independent techniques relative to Table~\ref{tab:distill}, but still outperforms the baselines reported in Table~\ref{tab:distill}. This suggests that teacher-side behavioral evidence is weakened but not removed by deployment-time changes. RAG is the most favorable setting, possibly because the retrieved context constrains the response space and makes teacher-student preference alignment easier to observe. By contrast, sampling strategies cause the largest degradation in TPR@1\%FPR, indicating that DistillTrace is particularly sensitive to sampling factors such as temperature under strict low-false-positive constraints.

\subsection{Ablation Study}

In this section, we investigate the hyperparameters that affect DistillTrace. Experiments on the hyperparameters of AttnPrint are provided in the supplementary material.

\begin{figure*}[t]
\centering
\begin{subfigure}[t]{0.24\textwidth}
\centering
\includegraphics[width=\textwidth]{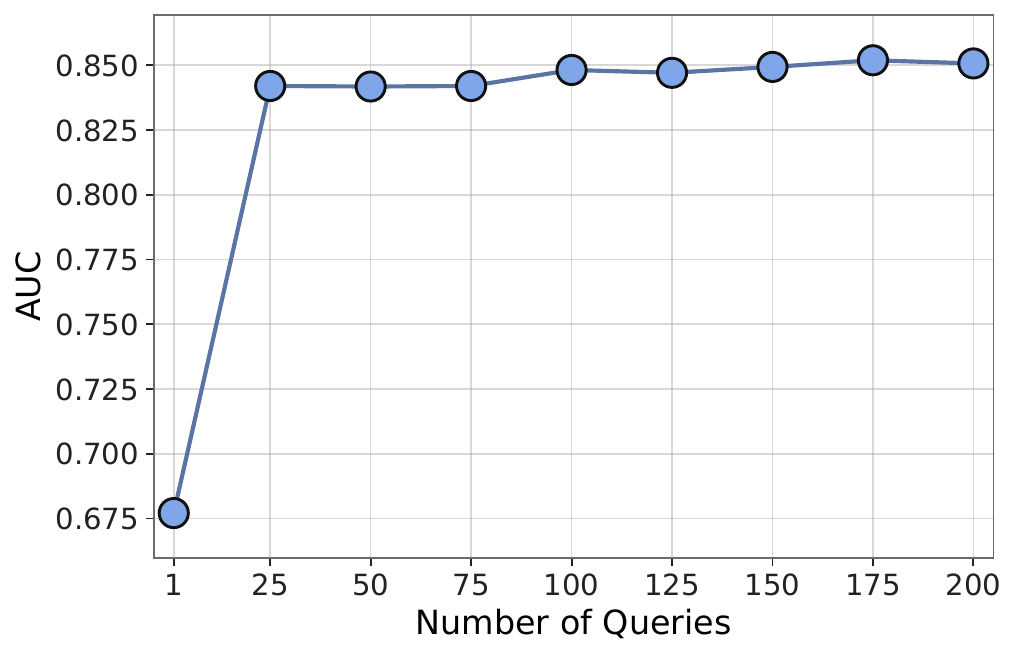}
\caption{Effect of query budget on AUC.}
\label{fig:ablation_query_auc}
\end{subfigure}
\hfill
\begin{subfigure}[t]{0.24\textwidth}
\centering
\includegraphics[width=\textwidth]{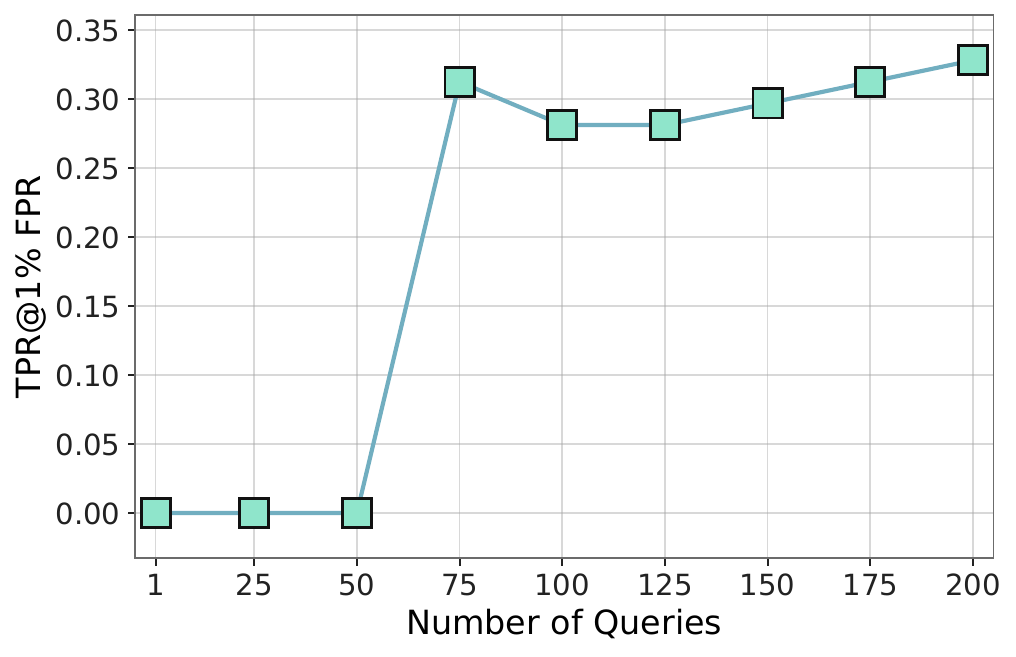}
\caption{Effect of query budget on TPR@1\%FPR.}
\label{fig:ablation_query_tpr}
\end{subfigure}
\hfill
\begin{subfigure}[t]{0.24\textwidth}
\centering
\includegraphics[width=\textwidth]{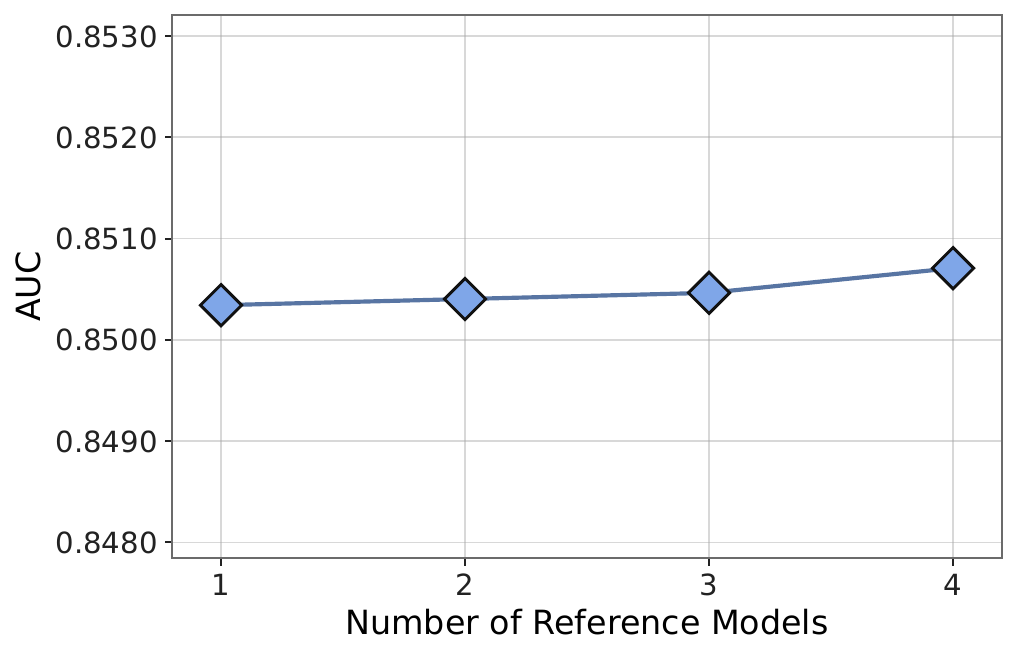}
\caption{Effect of reference model count on AUC.}
\label{fig:ablation_ref_auc}
\end{subfigure}
\hfill
\begin{subfigure}[t]{0.24\textwidth}
\centering
\includegraphics[width=\textwidth]{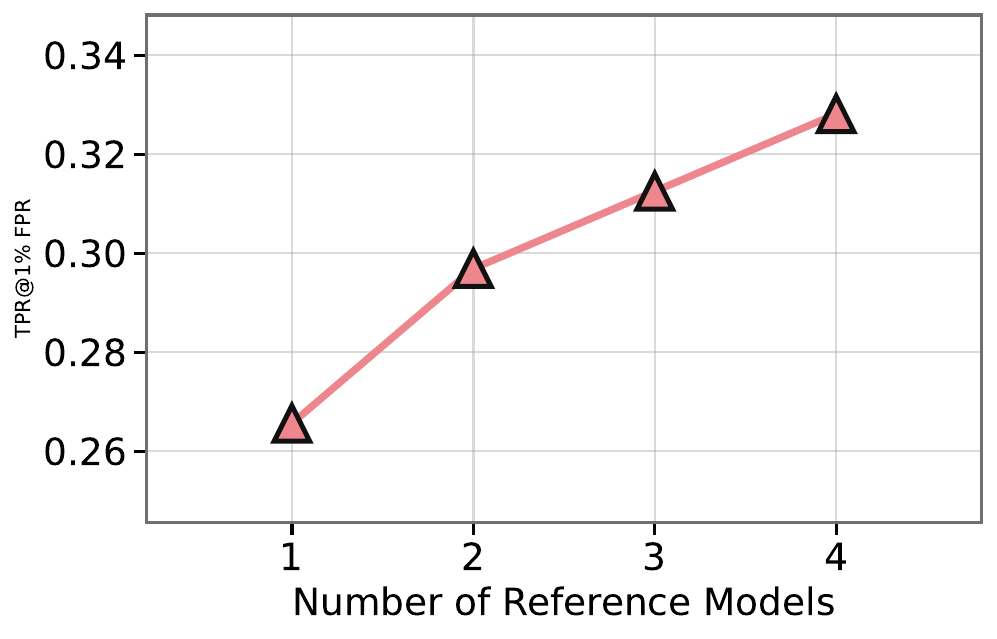}
\caption{Effect of reference model count on TPR@1\%FPR.}
\label{fig:ablation_ref_tpr}
\end{subfigure}
\caption{Detection performance of DistillTrace under different query budgets and reference model counts. Subfigures (a) and (b) illustrate the effect of the query budget, while (c) and (d) show the effect of the number of reference models.}

\Description{A four-panel figure showing the effect of query budget
and reference model count on DistillTrace. Panel (a) plots AUC
against query budget, panel (b) plots TPR@1\%FPR against query
budget, panel (c) plots AUC against reference model count, and panel
(d) plots TPR@1\%FPR against reference model count. AUC improves
quickly and then saturates as query budget increases, while TPR@1\%FPR is more sensitive to larger query budgets and also improves with
more reference models.}
\label{fig:ablation_panels}
\end{figure*}

\noindent \textbf{Effects of Different Numbers of Queries.} As shown in Figure~\ref{fig:ablation_query_auc} and Figure~\ref{fig:ablation_query_tpr}, increasing the query budget improves detection performance. AUC rises from 0.677 to 0.842 as the number of queries increases from 1 to 25, then gradually saturates around 0.85. TPR@1\%FPR is more sensitive and remains near zero with fewer than 50 queries, but improves substantially at 75 queries before stabilizing. Notably, such a query cost is acceptable in practice. For example, with 75 queries, where each query contains approximately 1600 image tokens, 33 input text tokens, and 128 output tokens, querying the suspect model costs approximately \$0.2121 under current mainstream commercial API pricing. Even with 200 queries, the total cost is only \$0.5655. These results suggest that our method is economically feasible in practical black-box auditing scenarios.

\noindent \textbf{Effect of the Number of Reference Models.} As shown in Figure~\ref{fig:ablation_ref_auc} and Figure~\ref{fig:ablation_ref_tpr}, the number of reference models has a relatively limited impact on the overall AUC. Even with only one reference model, DistillTrace already achieves a high AUC, indicating that it retains strong global discriminative ability even under extremely limited reference information. However, as the number of reference models increases, TPR@1\%FPR improves more substantially, rising from 0.2656 to 0.3281. This suggests that a larger reference pool provides a more stable estimate of the output distribution of independent models, allowing the student’s relative advantage over non-distilled models to be reflected more consistently in the detection statistic. Notably, even with only a single reference model, our method still outperforms existing model fingerprinting methods, further demonstrating the effectiveness and practicality of this black-box distillation detection framework in scenarios with limited reference resources.

\vspace{-\baselineskip}
\section{Conclusion}

This paper presents the first systematic study of model fingerprinting in the MLLM setting. Specifically, for model modification, we propose AttnPrint, a white-box method that identifies derivative relationships through stable low-frequency cross-modal attention features in the frequency domain. For unauthorized distillation, we propose DistillTrace, a black-box method that attributes distillation by comparing the confidence assigned by candidate teacher models to
suspect outputs against reference models and applying hypothesis testing. Large-scale experiments show that both methods consistently outperform existing approaches in identification accuracy and robustness across mainstream MLLM architectures and complex modification settings. Despite these promising results, our framework still has limitations. AttnPrint relies on white-box access to internal attention information, which may not always be available in real-world auditing. DistillTrace requires a task-relevant query set that can approximate the distillation task distribution, which also limits its applicability in real-world scenarios. Future work will explore MLLM fingerprinting under weaker access assumptions, especially black-box methods for model modification detection, and extend copyright auditing to broader model variants and deployment settings.

\begin{acks}
This work was supported in part by the New Generation Artificial Intelligence-National Science and Technology Major Project (No. 2025ZD0123202) and by National Natural Science Foundation of China (Grants U2336206 and 62472398).
\end{acks}

\bibliographystyle{ACM-Reference-Format}
\bibliography{ref}

\end{document}